\documentclass[Physsubmission, Phys]{SciPost}

\hypersetup{
    colorlinks,
    linkcolor={red!50!black},
    citecolor={blue!50!black},
    urlcolor={blue!80!black}
}
\usepackage[T1]{fontenc}
\usepackage[utf8]{inputenc}

\usepackage[bitstream-charter]{mathdesign}
\DeclareSymbolFont{usualmathcal}{OMS}{cmsy}{m}{n}
\DeclareSymbolFontAlphabet{\mathcal}{usualmathcal}

\begin{document}

\begin{center}{\Large \textbf{
Backward-Angle (u-Channel) Meson Production from JLab 12~GeV Hall C to EIC
\\
}}\end{center}

\begin{center}
W. B. Li\textsuperscript{1$\star$}
\end{center}

\begin{center}
{\bf 1} William \& Mary, Williamsburg VA 23185, USA
\\
* billlee@jlab.org
\end{center}

\begin{center}
\today
\end{center}


\definecolor{palegray}{gray}{0.95}
\begin{center}
\colorbox{palegray}{
  \begin{tabular}{rr}
  \begin{minipage}{0.1\textwidth}
    \includegraphics[width=22mm]{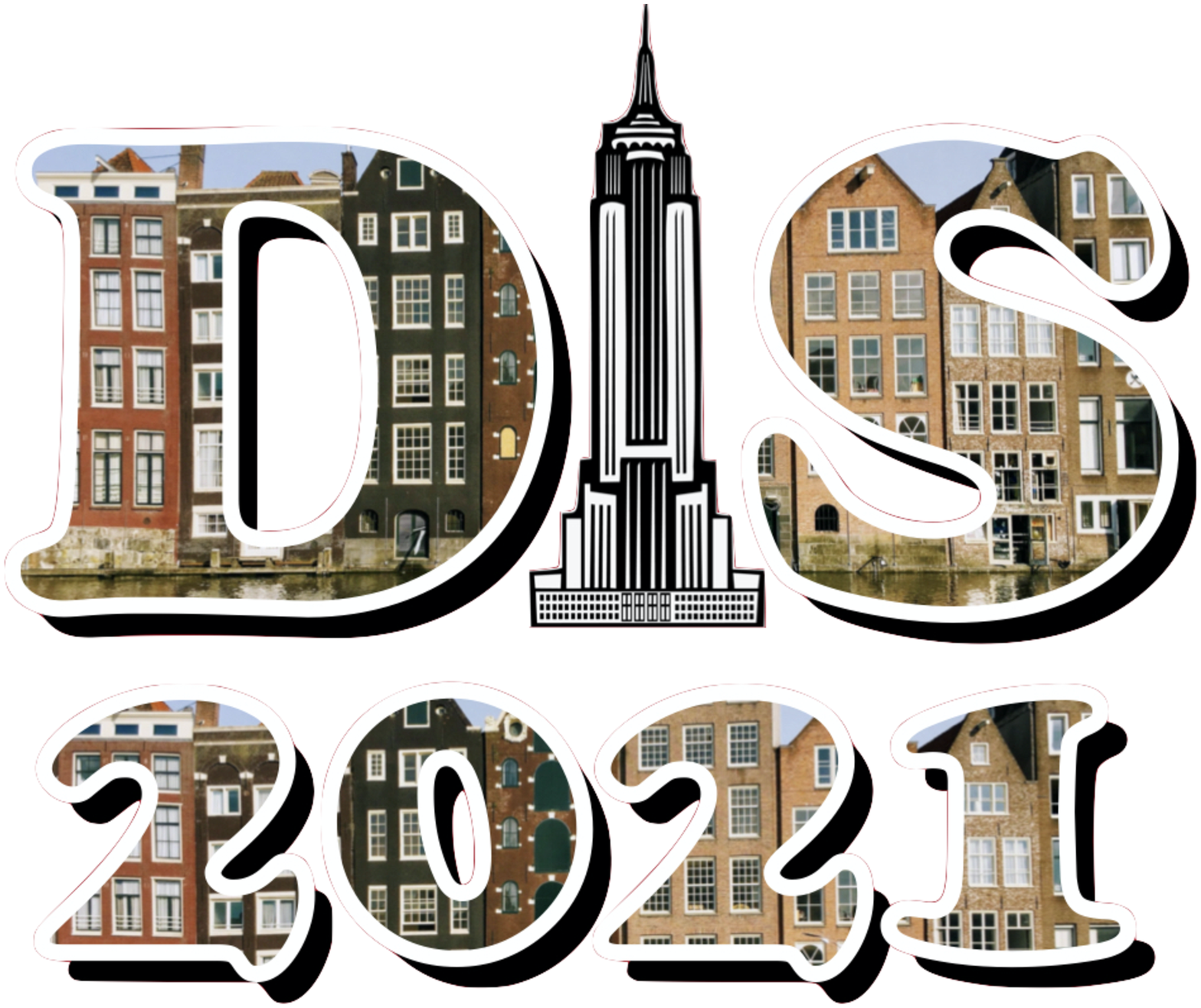}
  \end{minipage}
  &
  \begin{minipage}{0.75\textwidth}
    \begin{center}
    {\it Proceedings for the XXVIII International Workshop\\ on Deep-Inelastic Scattering and
Related Subjects,}\\
    {\it Stony Brook University, New York, USA, 12-16 April 2021} \\
    \doi{10.21468/SciPostPhysProc.?}\\
    \end{center}
  \end{minipage}
\end{tabular}
}
\end{center}

\section*{Abstract}
{\bf

The recent exclusive backward-angle electroproduction of $\omega$ from Jefferson Lab Hall~C electron-proton fixed-target scattering experiments above the resonance region hints at a new domain of applicability of QCD factorization in a unique $u$-channel kinematics regime. Thanks to this effort, the interest in studying nucleon structure through $u$-channel meson production observables has grown significantly. In the fixed target configuration, the $u$-channel meson electroproduction observables feature a unique interaction picture: the target proton absorbs nearly all momentum induced by virtual photons and recoils forward, while the produced mesons (such as omega or pions) are left behind almost at rest near the target station. In this presentation, We provide a summary of the key observations of the existing $u$-channel meson production results, update-to-date theory insights, and a path to further exploration from JLab 12 GeV Hall~C program to the future Electron-Ion Colliders. 
}


\vspace{10pt}
\noindent\rule{\textwidth}{1pt}
\tableofcontents\thispagestyle{fancy}
\noindent\rule{\textwidth}{1pt}
\vspace{10pt}

%
%

\section{Introduction}

Studying nucleon structure through probes of real and virtual photons has been a key objective of the hadron physics community in the past decades. 

\begin{figure}[t]
\centering
    \includegraphics[width=0.80\textwidth]{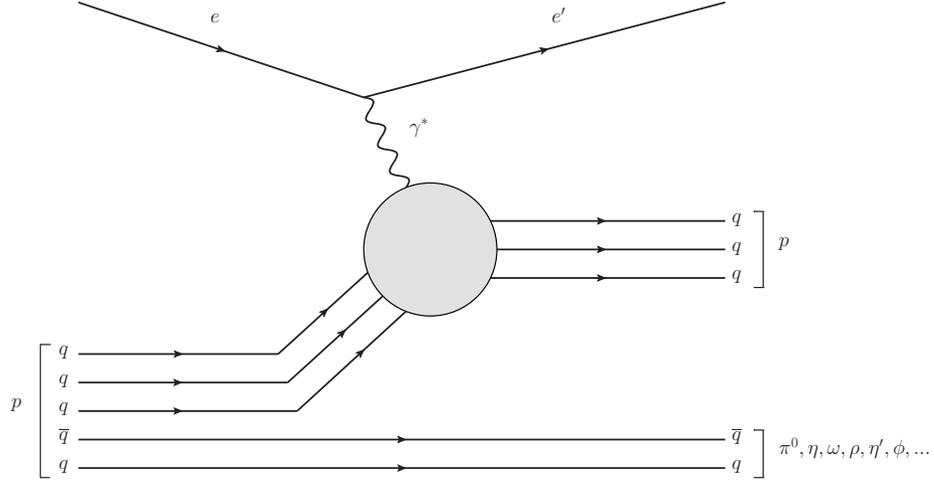}
    \caption{Illustration of $u$-channel (backward) meson production (a.k.a. knocking a proton out of a proton) process, includes: $\pi^0$, $\eta$, $\omega$, $\rho$, $\eta^\prime$, $\phi$, etc.}
\label{fig:proton_out} 
\end{figure}

Here, we focus on one specific type of interaction, known as the backward angle (or $u$-channel) exclusive interaction, where the squared momentum transfer $u$ between a produced meson $M$ and the target $N$ is such that $|u|=|(p_M-p_N)^2| \ll |t|=|(p_N^{\prime}-p_N)^2|$. The non-intuitive nature of such an interaction is revealed in the following example: a real ($\gamma$) or virtual photon ($\gamma^*$) probe is induced by the accelerated electron and interacts with a fixed proton target. The coiling nucleon absorbs most of the momentum transfer from the probe and travels forward, whereas the produced meson remains close to the target, nearly at rest. This type of reaction is sometimes referred to as a ``knocking a proton out of a proton'' process and offers improved access to the valence quark plus sea components of the nucleon wave function. Fig.~\ref{fig:proton_out} illustrates the non-intuitive nature of the $u$-channel production mechanism of mesons.

To ensure the final state dynamics are not dominated by resonance contributions, the invariant mass $W=\sqrt{p_{\gamma}^2+p_N^2}$ is chosen to be above the nucleon resonance region ($W>2$~GeV). In addition, all final state particles must be directly detected or indirectly reconstructed (using the missing mass reconstruction technique) to ensure exclusivity.

In these proceedings, the theoretical framework used to extract nucleon structure information is introduced; followed by a short overview of the current $u$-channel physics program at Jefferson Lab (JLab) 12 GeV, and future opportunities with the  Electron-Ion Collider (EIC)~\cite{AbdulKhalek:2021gbh} and Electron-Ion Collider in China (EIcC)~\cite{Anderle2021}.

\section{A factorization scheme in $u$-channel kinematics regime}
\label{sec:theory}

\begin{figure} 
\centering
    \includegraphics[width=0.49\textwidth]{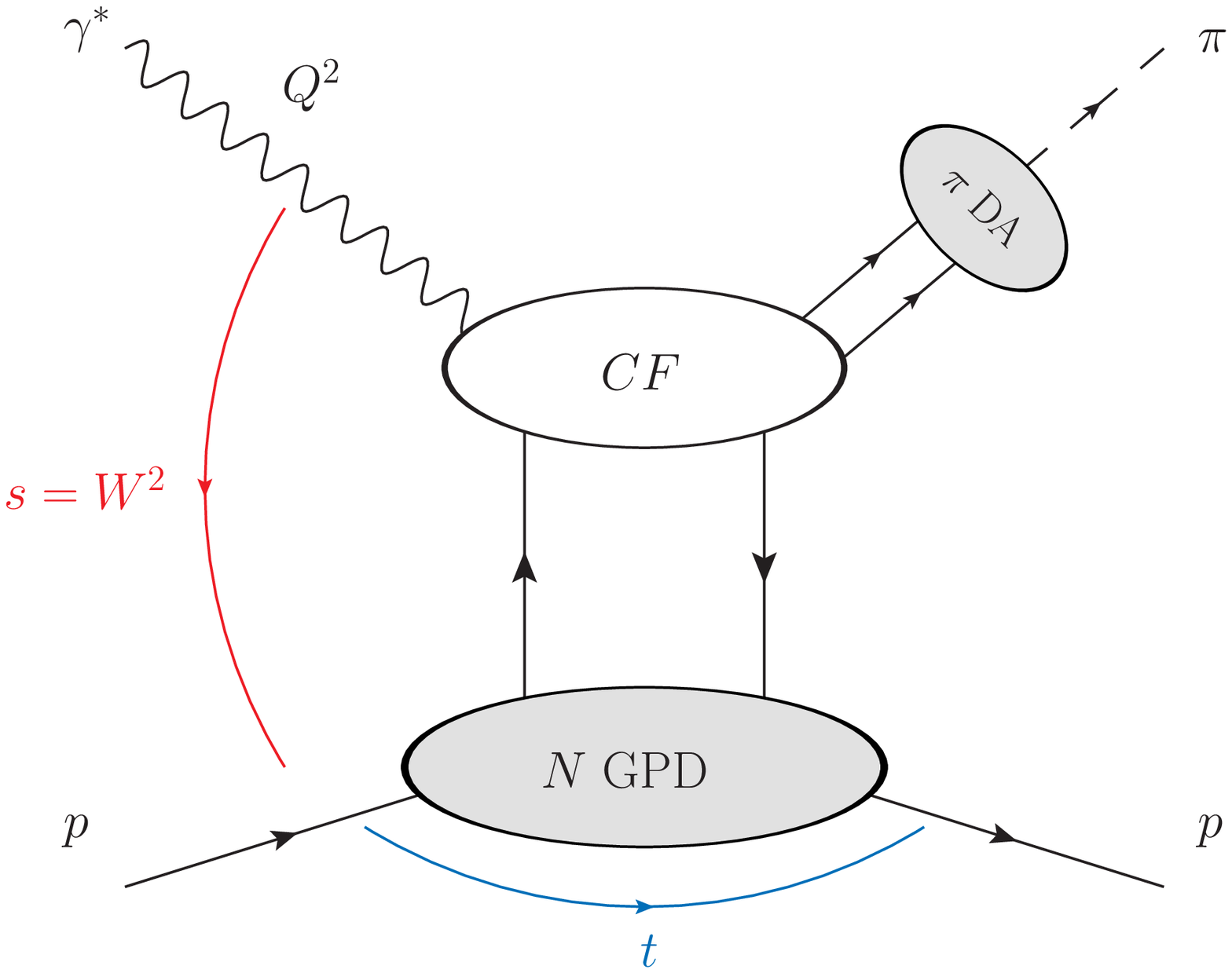}
    \includegraphics[width=0.49\textwidth]{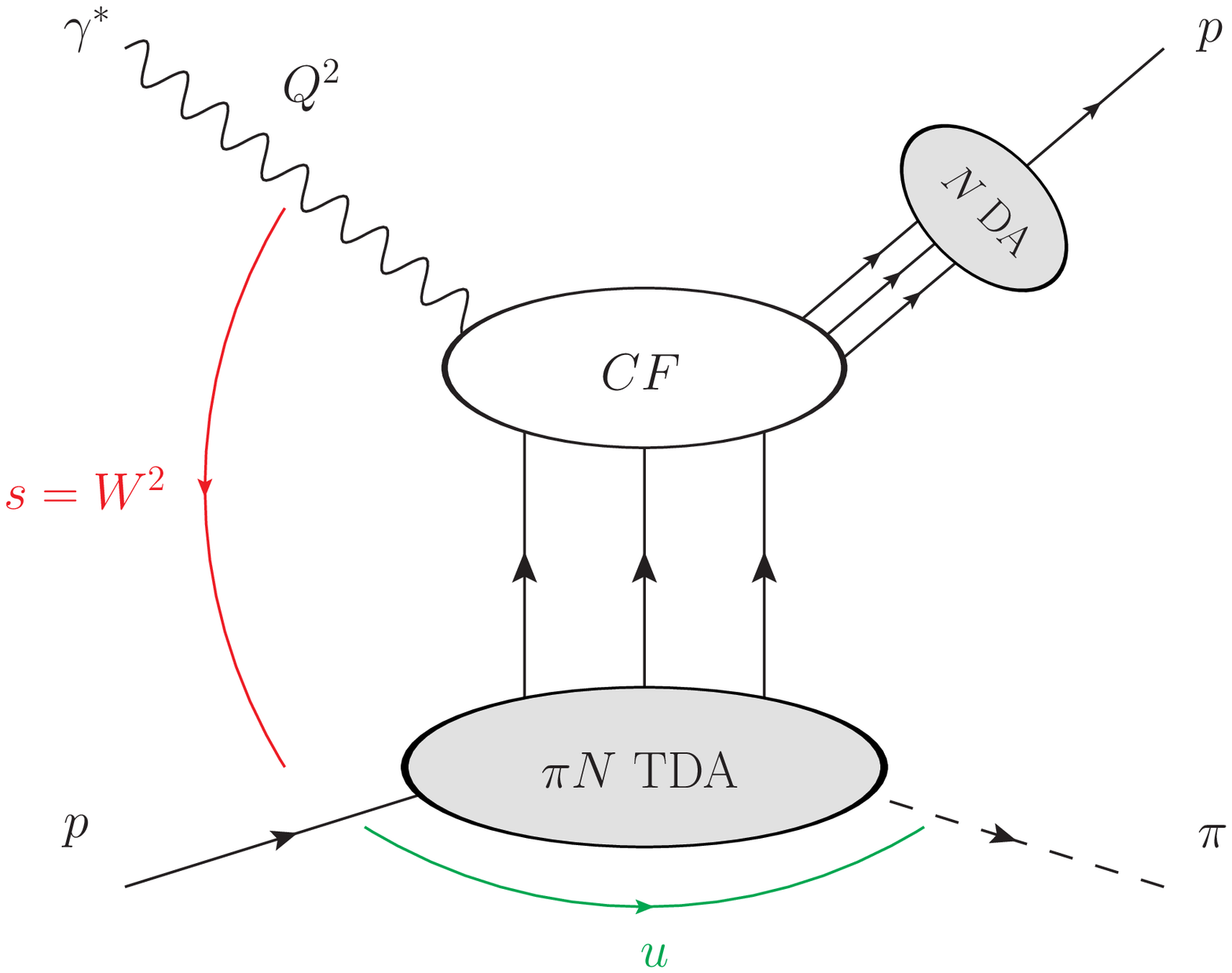}
    \caption{Left: shows the $\pi^0$ electroproduction interaction ($\gamma^*p\rightarrow p\pi^0$) diagram under the (forward-angle) GPD collinear factorization regime (large $Q^2$, large $s$, fixed $x_{\rm B}$, fixed $t\sim t_{min}$). $N$ GPD is the quark nucleon GPD (note that there are also gluon GPD that is not shown). $\pi$ DA stands for the vector meson distribution amplitude. The CF corresponds to the calculable hard process amplitude. Right: shows the (backward-angle) TDA collinear factorization regime (large $Q^2$, large $s$, fixed $x_{\rm B}$, $u\sim u_{min}$) for $\gamma^*p\rightarrow  p \pi^0$. The $\pi N$ TDA is the transition distribution amplitude from a nucleon to a vector meson.}
\label{fig:GPD_TDA} 
\end{figure}

The baryon-to-meson Transition Distributions Amplitudes (TDAs)~\cite{Pire:2010if,Pire:2011xv,Pire:2015kxa} are the backward-angle analog of Generalized Distribution Amplitudes~\cite{Ji:1997,Diehl:2003}.  TDAs describe the underlying physics mechanism of how the target proton transitions into a meson during the final state. One fundamental difference between GPDs and TDAs is that the TDAs require three parton exchanges between $\pi N$ TDA and Colinear Factorization (CF) in backward angle kinematics: $-t \rightarrow -t_{max}$, $-u \rightarrow -u_{min}$, $t>Q^2$ and $W>2$~GeV. The interaction diagrams of GPDs and TDAs are represented in Fig.~\ref{fig:GPD_TDA}.

The TDA colinear factorization has made two specific qualitative predictions regarding backward meson electroproduction, which can be verified experimentally~\cite{Lansberg:2011aa, Pire:2015iza, Pire:2015kxa, Pire:2019hos}:
\begin{itemize} 
\item The dominance of the transverse polarization of the virtual photon results in the suppression of the $\sigma_{\rm L}$ cross section by a least $1/Q^2$ : $\sigma_{\rm L}/\sigma_{\rm T} < 1/Q^2$ ,

\item The characteristic $1/Q^8$-scaling behavior of the transverse cross section for fixed $x$, following the quark counting rules.
\end{itemize}
These predictions were validated by the cross section measurement of the exclusive backward $\pi^+$ electroproduction from CLAS~\cite{Park:2017irz} and the L/T separated $\omega$ cross section from Hall C~\cite{Li:2019xyp} (both were JLab 6~GeV data). Based on these initial successes, a program to systematically study TDAs for a given reaction requires the following three stages:
\begin{description}
\item[Stage I:] Further validating of the TDA framework, by measuring the general scaling trend of the L/T cross section ratios for $\pi^0$, $\eta$, $\rho$, $\omega$, $\eta^{\prime}$, $\phi$, and other meson production channels.
\item[Stage II:] Determination of the $-u$ dependence of the cross section and extraction of the meson to nucleon transition form factors.
\item[Stage III:] Extraction the TDAs by probing the single and double spin asymmetries for backward meson production at 12~GeV JLab programs and EIC.
\end{description}


\section{Current exclusive $\pi^0$ production programs at JLab 12 GeV}

\begin{figure} 
\centering
  \includegraphics[width=0.75\textwidth]{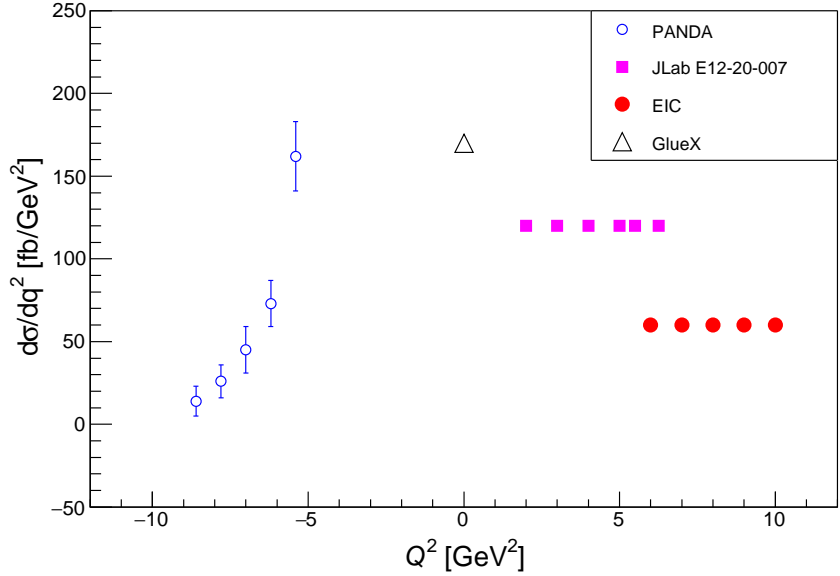}
\caption{The anticipated global data set of $d\sigma/dq^2 (\gamma^*p \rightarrow p\pi^0)$ vs $Q^2$ at fixed $s=10$ GeV$^2$. Projected results from $\overline{\textrm{P}}$ANDA (FAIR) $q^2 = - Q^2$ scaling are in open blue circle; projected JLab E12-20-007 measurements are in magenta square; projected EIC measurements are in red full circle; JLab GlueX photoproduction measurement ($Q^2 = 0$ GeV$^2$) is indicated by the open triangle. These experimental programs are elaborated in the relevant subsections.}
\label{fig:u_channel_pi0} 
\end{figure}

Similar to the 6~GeV era Hall~C measurements, additional $u$-channel meson electroproduction data were fortuitously acquired during the KaonLT experiment (E12-09-011) \cite{E12-09-011}.  The primary purpose for the acquisition of these data was the study of the $K^+$ electromagnetic form factor, but the detector apparatus allowed $^1$H$(e,e'p)X$ data to be acquired in parallel.  Data were taken well above the resonance region ($W=2.32$--3.02 GeV), at selected settings between $Q^2=0.5$ and 5.50 GeV$^2$.  For each $Q^2, W$ setting, data were taken at two beam energies, corresponding to $\Delta\epsilon\sim 0.25$, so that L/T/LT/TT separations could be performed.

The recently approved JLab E12-20-007~\cite{E12-20-007} is the first dedicated $u$-channel physics experiment and is an important small step towards the objectives to study TDAs (\textbf{Stage I} and \textbf{II} from Sec.~\ref{sec:theory}). The measurement aims to probe the $^1$H$(e, e^{\prime}p)\pi^0$ exclusive electroproduction reaction over the $2<Q^2<6.25$~GeV$^2$ kinematic range, at fixed $W=3.1$~GeV ($s=10$~GeV$^2$) and $-u_{min}$. The experiment will utilize the 11 GeV $e$ beam on an unpolarized liquid hydrogen target (LH$_2$), in combination with the high precision High Momentum Spectrometer (HMS), Super High Momentum Spectrometer (SHMS) available in Hall~C. The key observable involves the detection of the scattered electrons in coincidence with energetic recoiled protons, and resolving the exclusive $\pi^0$ events using the missing mass reconstruction technique~\cite{workshop2021}. The separated cross sections, $\sigma_{T}$, $\sigma_{L}$, and the $\sigma_{T}/\sigma_{L}$ ratio at 2-5 GeV$^2$, will directly challenge the two predictions of the TDA model, $\sigma_T = 1/Q^8$ and $\sigma_T \gg \sigma_L$, in $u$-channel kinematics. This will be an important step forwards validating the existence of a backward factorization scheme and establishing its applicable kinematics range.


A preliminary study~\cite{AbdulKhalek:2021gbh} has confirmed the feasibility of studying $e+p\rightarrow e'+p'+\pi^0$ over the range $6.25 < Q^2 < 10.0$~GeV$^2$. The EIC offers a unique opportunity to provide a definitive test of TDA predictions beyond JLab 12 GeV kinematic (see Sec.~\ref{sec:future}). Backward $\pi^0$ production will be studied by the $\overline{\textrm P}$ANDA experiment at FAIR~\cite{Singh:2014pfv,Singh:2016qjg}. This experimental channel can be accessed through observables including $\overline{p} + p \rightarrow \gamma^* + \pi^{0}$ and $\overline{p} + p \rightarrow J/\psi + \pi^0$ . Note that this backward $\pi^0$ production involves the same TDAs as in the electroproduction case. They will serve as very strong tests of the universality of TDAs in different processes~\cite{Singh:2014pfv,Singh:2016qjg}. Fig.~\ref{fig:u_channel_pi0} underlines a systematic program in studying $Q^2$ evolution of the exclusive $\pi^0$ production in $-10<Q^2<10$~GeV$^2$, at fixed $W\sim 3.15$~GeV ($s=10$~GeV$^2$) and $-u\sim u_{min}$. Data sets from JLab E12-20-008, EIC and $\overline{\textrm{P}}$ANDA will play their unique roles to cover different kinematics regions.

\section{Future $\pi^0$ Production at EIC and EICC}
\label{sec:future}
The Electron-Ion Collider (EIC) is a new, innovative, large-scale particle accelerator facility conceived by the nuclear physics community over two decades and it's planned for construction at Brookhaven National Laboratory. EIC provides $e+p$ center-of-mass energy from 20-100~GeV at a high collision luminosity of $10^{33}$-$10^{34}$ cm$^{-2}$s$^{-1}$. The large acceptance and forward/backward tagging capabilities make the EIC a perfect ground to extensively study the $u$-channel meson production processes. 

In the $\pi^0$ electroproduction sector, the impact of EIC data is illustrated in Fig.~\ref{fig:u_channel_pi0}.  It shows a prospective $Q^2$ ($10 < Q^2 < 10$ GeV$^2$) evolution, combining backward ($u \sim u_{\textrm {min}}$) exclusive $\pi^0$ production data from JLab E12-20-007, $\overline{\textrm{P}}$ANDA, and EIC, at fixed $s=10$ GeV$^2$. 
A preliminary study has confirmed the feasibility of studying the  $e+p\rightarrow e'+p'+\pi^0$ interaction in the range: $6.0 < Q^2 < 10.0$~GeV$^2$. A data set combining E12-20-007 and EIC for exclusive $\pi^0$ production will offer a definitive challenge to the $1/Q^{8}$ scaling prediction of the TDA formalism.


\begin{figure} 
\centering
    \includegraphics[width=1\textwidth]{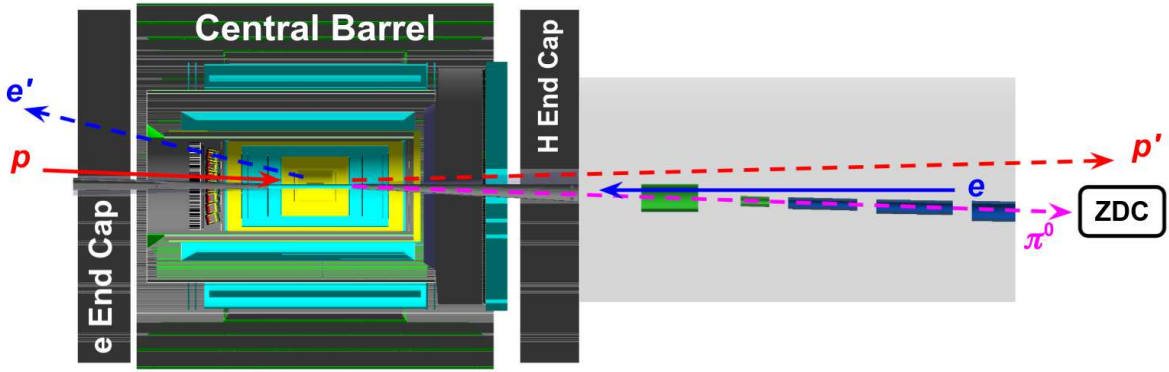}
    \caption{Schematic display of a typical $u$-Channel $\pi^0$ electroproduction event at EIC. Solid lines represent the beam particles and dashed lines represent the final state particles.}
\label{fig:u_channel_pi0_event} 
\end{figure}

\begin{figure} 
\centering
  \includegraphics[width=0.7\textwidth]{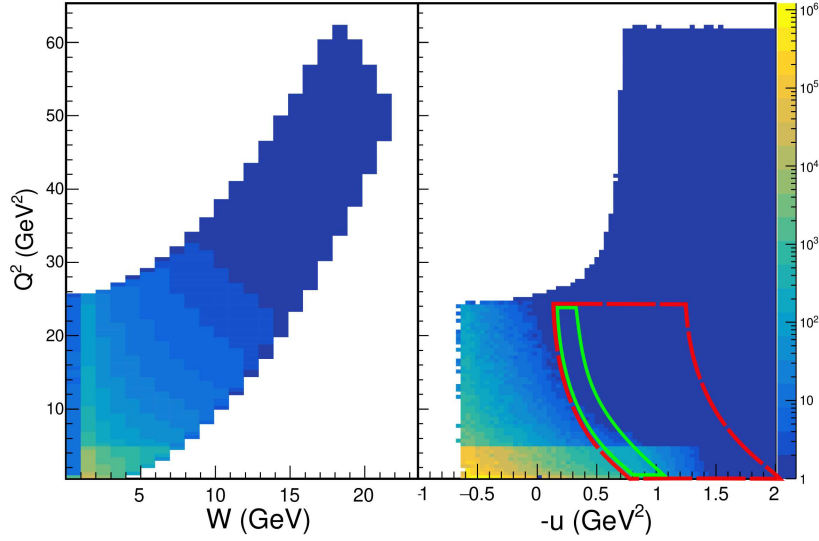}
  \caption{Left: $Q^2$ vs $W$ coverage for $e+p\rightarrow e^\prime+ p^\prime + \pi^0$, where at least one $\gamma$ is detected by the ZDC. Right: $Q^2$ vs $-u$ coverage for all available $s$ values, $0<s<400$ GeV$^2$. The red enclosed region represents $Q^2$ vs $-u$ coverage at $9<s<11$ GeV$^2$; the green enclosed region is a subset of the red, and presents the coverage of events with both photons detected by the Zero-Degree Calorimeter.} 
\label{fig:u_channel_phasespace} 
\end{figure}

The study of $e+p\rightarrow e^\prime+ p^\prime + \pi^0$ at $-u_{min}$ arises surprisingly naturally, thanks to the 4$\pi$ coverage of the EIC detector package and forward-tagging capability. A feasibility study has shown the optimal collision energy option: 5 GeV electron beam on a 100 GeV proton~\cite{AbdulKhalek:2021gbh}, for fixed $s=10$ GeV$^2$ at $u\sim u_{min}$. The corresponding available $Q^2$ vs $-u$ phasespace is shown in the right panel of Fig.~\ref{fig:u_channel_phasespace}.
At the kinematic range of interest, the scattered electrons at pseudorapidity $|\eta|<1.5$ and $p_{e}\sim 5.4$~GeV/c will be detected by the Electron-End-Cap, well within the EIC specification
~\cite{Accardi:2012qut, EIC:RDHandbook, AbdulKhalek:2021gbh}; the Zero Degree Calorimeter (ZDC) will be used to detect decayed $\pi^0\rightarrow\gamma\gamma$ for momenta from 40 to 60 GeV/c; for the forward recoiled proton, the detector and material studies show the Hadron-End-Cap will provide a silicon tracker to cover $\eta$ range up to 3.5: $|\eta|<3.5$. One would need a dedicated detector to tag the recoiled proton at $\eta \sim 4.1$ at $\phi = 180^\circ$, otherwise, the missing mass reconstruction technique will be applied to resolve the proton. Note that the feasibility of the missing mass reconstruction technique remains to be demonstrated at the EIC.

To extract the differential cross section of the exclusive $\pi^0$ events, the event selections include the following scenarios:
\begin{itemize}

\item{All final state particles are detected, including $e^{\prime}$, $p^{\prime}$ and 2$\gamma$. A feasibility study~\cite{AbdulKhalek:2021gbh} projected 20 to 30\% double $\gamma$ detection efficiency for $\pi^0$ at 40 to 60 GeV/c, respectively. Here, $p_\pi=$40 GeV/c corresponds to $Q^2 \sim 10$ GeV$^2$.  It is also worth noting the hit pattern of the two photons forms a ring around the high occupancy spectator neutron region at the ZDC plane from other tagged diffractive processes.} 

\item{$e^\prime$, $p^\prime$ and a single $\gamma$ (from decayed $\pi^0$) are detected. The lost photon will likely be consumed by the steering magnet arrays upstream of the ZDC. Under this scenario, one could rely on a detailed simulation of the known physics backgrounds, such as $u$-channel DVCS, $\eta\rightarrow2\gamma$ and $\omega\rightarrow\pi^0\gamma$, in addition to the relative normalization of the expected 2$\gamma$ efficiency (from scenario 1) to extract the yield.}

\item{$e^\prime$ and 2$\gamma$ are detected. Although there are ongoing experimental efforts to ensure the detection of the forward recoiling proton, there is a small possibility the proton signal is rejected as background, which will complicate this scenario.  Here, the coplanarity of the two $\gamma$ that hit the ZDC will play a significant role in identifying $\pi^0$ events, and the reconstructed massing mass distribution may resolve the missing proton at the desired kinematics setting.} 

\end{itemize}

As a source of physics background, the 3$\gamma$ final state through decay mode $\omega\rightarrow \pi^0\gamma$ has a branching ratio of 8.28\%~\cite{PDG2020}. Although it is possible for $\omega\rightarrow \pi^0\gamma$ to contaminate the exclusive $\pi^0$ event sample in all three trigger scenarios, it is possible to minimize this effect: examine angular coplanarity (back-to-back) in the center-of-mass frame for the two $\gamma$ that hit the ZDC; initiate a boundary in the missing mass distributions to exclude $\omega$ events. A full simulation study should give further insight on the level of experimental background and the most effective methodologies for removing them.

With a lower center-of-mass energies from 15 to 20 GeV, the Electron-Ion Collider in China (EicC) offers unique opportunities to carry out $e+p$ studies to fill the small gap in kinematics coverage of JLab 12 GeV and EIC. The nominal 20 GeV proton beam colliding with a 3.5 GEV electron beam will produce $u$-channel $\pi^0$ at much lower momentum ($p_{\pi^0} < 40$ GeV). The decayed photons (from $\pi^0$ decay at $u\sim u_{min}$) will require a unrealistically large ($>15$~mRad) ZDC acceptance for the simultaneous detection. On the other hand, the lower beam proton momentum reduces the Lorentz boost in the proton beam momentum, thus, significantly extends the accessible $u$-coverage (for detecting $\gamma$s) and reduced the `dead zone' caused by the beamline components (in the forward forward region) at $|-u|>|-u_{min} + 0.5|$. There is great possibility to access the full $u$ coverage using the combined EIC+EIcC converge for the overlapped kinematics region, and this requires detailed simulation studies to achieve.

\bibliography{refs}

%
%
%
%
%
%
%
%

\end{document}